\documentstyle[aps,preprint]{revtex}

\begin{document}
\draft

\title{Measurement of Superluminal optical tunneling times in double-barrier
photonic bandgaps}
\normalsize
\author{S. Longhi and P. Laporta} 
\address{Istituto Nazionale di Fisica per la Materia, Dipartimento di Fisica, Politecnico di Milano, Piazza L. da
Vinci 32,  I-20133 Milan, Italy}
\author{M. Belmonte} 
\address{Corning-Optical Technologies Italia S.p.A., V.le Sarca 222, 20126
Milan, Italy}
\author{E. Recami} 
\address{Facolt\`a di Ingegneria, Universit\`a statale di Bergamo, Dalmine (BG), Italy;}
\address{INFN--Sezione di Milano, Milan, Italy; {\rm and}}
\address{C.C.S., UNICAMP, Campinas, S.P., Brasil.}
\maketitle
\tightenlines

\small
\begin{abstract}
Tunneling of optical pulses at 1.5 $\mu$m wavelength through double-barrier
periodic fiber Bragg  gratings is experimentally investigated. Tunneling time
measurements as a function of barrier distance show that, far from the
resonances of the structure, the transit time is paradoxically short,
implying Superluminal propagation, and almost independent of the distance
between the barriers. These results are in agreement with theoretical
predictions based on phase time analysis and also provide an experimental
evidence, in the optical context, of the analogous phenomenon expected
in Quantum Mechanics for non-resonant Superluminal tunneling of particles
across two successive potential barriers. [Attention is called, in particular,
to our last Figure]. \\ 
\end{abstract}
\newpage
\section{Introduction}
Tunneling of a particle through a potential barrier is one of the most
intriguing phenomena in quantum mechanics, 
that continues to attract a great attention both theoretically and
experimentally. The reason thereof stems from the fact that  some  questions
related to the dynamics of tunneling, such as the definition and measure of
tunneling transit times \cite{Hauge89,Olkh92,Landauer94}, have not get a general acceptance yet.
In addition, tunneling hides some amazing time-domain phenomena, the most
notable one being Superluminal propagation (for related experiments see, e.g.,
\cite{Chiao97,Nimtz97} and references therein). In case of opaque barriers,
it is known that the tunneling time becomes independent of the barrier width
(Hartman effect \cite{Hartman62}; see also \cite{Olkh92}) and can become so short to imply
apparent Superluminal motion, which has been the subject of a lively debate
in recent years \cite{Chiao97,Nimtz97,Recami00,Pablo00}. Since tunneling time
measurements for
electrons are usually difficult to achieve and of uncertain interpretation,
experimental validation of the Hartman effect and direct measurements of
Superluminal tunneling times have been successfully obtained for the
closely-related problem of tunneling of photons through photonic barriers in
a series of famous experiments performed at either microwave
\cite{Enders93,Ranfagni93} and optical wavelengths
\cite{Steinberg93,Spielmann94,Balcou97}. In particular, in
Refs.\cite{Steinberg93,Spielmann94} one-dimensional photonic bandgaps were
used as photonic barriers, realizing the optical analogue of electron Bragg
scattering in the 
Kr\"{o}nig-Penney model of solid-state physics. A different but related issue is
that of particle tunneling through a double-barrier (DB) potential
structure, such as electron tunneling in semiconductor superlattice
structures. In this case, the resonant behavior of tunneling escape versus
energy \cite{Merzbacher70} is a clear manifestation of the wave nature of
electrons and  is of major importance for ultrahigh-speed resonant-tunneling
devices \cite{Capasso86}. Exploiting the analogy between electron and photon
tunneling \cite{Chiao91,Hupert77}, resonant tunneling phenomena have been also
studied and observed in connection with microwave propagation in undersized
waveguides and in periodic layered structures
\cite{Pablo00,Enders93,Cuevas96}, and general relations have been derived
between the traversal time at resonances and the lifetime of the resonant
states \cite{Cuevas96}. Besides of realizing resonance tunneling, it has
been theoretically recognized that DB structures are also of interest to
study {\em off-resonance} tunneling times
\cite{Pablo00,Enders93,Cuevas96,Olkhovsky01}. In this case, for opaque barriers it 
turns out that the transit
time to traverse the DB structure is independent not only of the barrier
width, but {\em even} of the length of the intermediate
(classically allowed) region that separates the barriers.
So far, measurements of Superluminal tunneling in DB structures
have been performed by G. Nimtz and coworkers in a series of microwave
transmission experiments \cite{Enders93}; however, no experimental
study on off-resonant tunneling times in DB photonic structures at optical wavelengths has been
reported yet. In recent works \cite{Longhi01,Longhi01b}, some of the 
present authors have shown that fiber Bragg gratings (FBGs) can provide versatile tools for the 
study of tunneling phenomena. Besides for their potential relevance in applications to optical communications, 
the use of FBGs as photonic barriers  is very appealing from an experimental viewpoint because the tunneling times 
in FBG structures fall in the tens of picoseconds time scale, which can be easily and precisely detected by standard optoelectronic 
means.\\ 
In this work we report on measurement of tunneling delay times in DB
photonic structures at the 1.5 $\mu$m  wavelength of optical communications. 
Our results represent an extension at optical wavelenghts of similar experimental achievements 
previously reported at microwaves \cite{Enders93} and provide 
 a clear experimental evidence that, for opaque barriers, the
traversal time is independent of barrier distance ({\em generalized} Hartman
effect). The paper is organized as follows. In Sec.II, the basic model of tunneling in a 
DB rectangular FBG is reviewed and the quantum-mechanical analogy of electron tunneling is outlined. 
In Sec.III the experimental measurements of tunneling times are presented. Finally, in Sec.IV the main conclusions are outlined.
\section{Optical tunneling in a DB FBG: basic equations and quantum-mechanical analogy}
We consider tunneling of optical pulses through a DB photonic structure achieved in a monomode optical fiber by writing
onto it two periodic Bragg gratings, each of width $L_0$, separated
by a distance $L$, which realize a weak modulation of the refractive index
$n$ along the  fiber axis $z$ according to $n(z)=n_0 \left[ 1+2 V(z) \cos(2
\pi z / \Lambda  ) \right ]$, 
where $n_0$ is the average refractive index of the structure, $\Lambda$ is
the Bragg modulation period, and $V(z)$ is profiled to simulate a symmetric
rectangular DB structure, i.e. $V(z)=V_0$ constant for $0<z<L_0$ and for
$L+L_0<z<L+2L_0$, and $V(z)=0$ otherwise (see Fig.1). For such a structure, Bragg
scattering of counterpropagating waves at a frequency $\omega$ close to the
Bragg resonance $\omega_B \equiv c_0 \pi / (n_0 \Lambda)$ ($c_0$ is the
speed of light in vacuum) occurs in the grating regions, whereas multiple
wave interference between the two barriers leads to Fabry-Perot resonances
in the transmission spectrum. 
The tunneling problem in the DB FBG structure bears a close connection to
that of non-relativistic electrons through a symmetric rectangular DB
potential, which has been widely investigated in literature (see, for instance,
\cite{Merzbacher70,Headings63}). The analogy is
summarized in Table I, where the basic equations and the expressions for
barrier transmission and group delay are given in the two
cases \cite{Hupert77,note0}.
 In the electromagnetic case, a monochromatic field
$E(z,t)$ at an optical frequency $\omega$ close to the Bragg frequency
$\omega_B$ propagating inside the fiber can be written as a superposition of
counter-propagating waves, $E(z,t)=u(z) \exp(-i \omega t+ik_Bz)+v(z) \exp(-i
\omega t -ik_Bz)+c.c.$, where $k_B= \pi /\Lambda$ is the Bragg wavenumber;
for a small index modulation ($V_0 \ll 1$), the envelopes $u,v$ of
counterpropagating waves satisfy  the following coupled-mode equations \cite{Ergodan97}:
\begin{mathletters}
\begin{eqnarray}
\frac{du}{dz} & = & i \delta u +i k_B V(z) u \\
\frac{dv}{dz} & = &- i \delta v -i k_B V(z) v
\end{eqnarray}
\end{mathletters}
where  $\delta \equiv
k-k_B=n_0(\omega-\omega_B)/c_0$ 
 is the detuning parameter between wavenumber $k= n_0 \omega / c_0$ of
counterpropagating waves and Bragg wavenumber $k_B$. The wave envelopes $u$ and $v$
are oscillatory (propagative) in the
region $L_0<z<L_0+L$, whereas they are exponential (evanescent) inside the
gratings when $| \delta| < k_B V_0$.  The spectral
transmission of the structure, given by $t(\omega)=[u(L)/u(0)]_{v(L)=0}$,
can be analytically determined by standard transfer matrix methods \cite{Ergodan97}.
As an estimate of  the tunneling time for a wavepacket crossing the
structure, we use the group delay (or phase time) as calculated by the
method of stationary phase, which is given by \cite{note1} $\tau = {\rm Im} \left\{
\partial {\rm ln}(t)  / \partial \omega \right\}$.  A typical
behavior of power transmission $T=|t|^2$ and group delay 
$\tau$ versus frequency detuning $\nu=(\omega-\omega_B)/(2 \pi)$, computed
for one of the DB structures used in the experiments, is shown in Fig.2.
Notice that, far from the sharp Fabry-Perot resonances, the group delay is
shorter than that for free propagation from input to
output planes, implying Superluminal propagation. At the center of the
bandgap ($\delta=0$), simple analytical expressions for the power transmission
and group delay can be derived and read:
\begin{equation}
T=\frac{1}{\cosh^2(2k_BV_0L_0)}
\end{equation}
\begin{equation}
\tau=\tau_1+\tau_2
\end{equation}
where:
\begin{mathletters}
\begin{eqnarray}
\tau_1 & = & \frac{n_0}{c_0 k_B V_0} {\rm tanh}(2k_B V_0 L_0) = \sqrt{1-T} \frac{n_0}{c_0 k_B V_0}\\
\tau_2 & = & \frac{n_0 L}{c_0} \frac{1}{{\rm cosh}(2k_B V_0 L_0)}= \sqrt T \frac{n_0 L}{c_0} 
\end{eqnarray}
\end{mathletters}
Equations (3) and (4) clearly show that two distinct contributions 
are involved in the expression for the group delay. The former term,
$\tau_1$, is independent of the barrier 
separation, and coincides with the tunneling time of a {\em single} barrier
of width $2L_0$. For an opaque barrier ($L_0V_0k_B \gg 1$), $\tau_1$ 
becomes independent of barrier width and saturates to the value $\tau_1 \sim
n_0/(c_0 k_B V_0)$ (Hartman effect). Conversely, the latter contribution,
$\tau_2$, is always shorter than the free-propagation time over a length $L$
and goes to zero for an opaque barrier, implying that the tunneling time
becomes independent of barrier distance ({\em generalized} Hartman effect).
Similar results are obtained for off-resonance tunneling of a
non-relativistic electron through a rectangular DB potential $V(z)$ assuming
that the incident wavepacket has a below-barrier mean energy $E$ half the
barrier width $V_0$; the corresponding expressions for barrier transmission and group delay in this case 
are given in Table I (for details, see, e.g., \cite{Headings63}). The tunneling through a DB FBG structure 
can hence be used as an experimental verifiable model for the quantum-mechanical case.\\ 
\section{Tunneling time measurements}
We performed a series of tunneling time measurements through DB FBG structures
operating at around 1.5 $\mu$m in order to assess the independence of
the peak pulse transit times with barrier distance $L$. The FBGs used for the
experiments were manufactured by using standard writing
techniques, with an exposure time to UV laser beam and 
phase mask length such as to realize a grating  with sharp fall-off edges of
length $L_0 \simeq 8.5$ mm and with a refractive index modulation $V_0
\simeq 0.9 \times 10^{-4}$. For such a refractive index modulation, a
minimum power transmission $T \simeq 0.8 \%$ at antiresonance is achieved
for a DB structure, which is low enough to get the opaque barrier limit but
yet large enough to perform time delay measurements at reasonable power
levels. The period of the phase mask was chosen to achieve Bragg resonance
at around 1550 nm wavelength.  Five different 
DB structures were realized with grating separation $L$ of 18, 27, 35, 42
and 47 mm. For such structures, both transmission spectra and group delays 
were measured using a phase shift technique \cite{Ryu89} with a spectral
resolution of $\simeq 2$ pm; an example of measured transmission spectrum
and group delay versus frequency for the 42-mm separation DB FBG is shown in
Fig.2. Notice that, according to the theoretical curve shown in the same
figure, far from the Fabry-Perot resonances the group delay is Superluminal,
with expected time advancement of the order of 240--250 ps. Notice also that
the sharp Fabry-Perot resonances are not fully resolved in the experimental
curves due to bandwidth limitations 
($\sim 2$ pm) of the measurement apparatus.\\
Direct time-domain measurements
of tunneling delay times were performed in transmission experiments using
probing optical pulse with $\simeq $1.3  ns duration, corresponding to a
spectral pulse bandwidth which is less than the frequency separation of
Fabry-Perot resonances for all the five DB structures. Since both
transmission and group delay are slowly varying functions of frequency far
from Fabry-Perot resonances (see Fig.2), weak pulse distortions are thus
expected for off-resonance pulse transmission. The experimental set-up for
delay time measurements is shown in Fig.3. A pulse train, at a repetition
frequency $f_m=300$ MHz, was generated by external modulation of a
single-frequency continuous-wave tunable laser diode  (Santec mod.
ECL-200/210), equipped with both a coarse and a fine (thermal) tuning control
of frequency emission with a resolution of $\sim 100$ MHz. The
fiber-coupled $\sim 10$ mW output power emitted by the laser diode was
amplified using a high-power erbium-doped fiber amplifier
(IPG Mod. EAD-2-PM; EDFA1 in Fig.3), and then sent to a ${\rm
LiNbO}_3$-based Mach-Zehnder modulator, sinusoidally driven at a frequency
$f_m=300$ MHz 
by a low-noise radio-frequency (RF) synthetizer.
The bias point of the modulator and the RF modulation power level
were chosen to generate a train with a pulse duration (FWHM) of $\simeq$~1.3
ns; the measured average output power of the pulse train available
for the transmission experiments was 
$\sim$ 130 mW. The pulse train was sent to the DB FBG through a three-port
optical circulator, that enables both transmitted and reflected signals to
be simultaneously detected. The signal transmitted through the DB FBG was
sent to a low-noise erbium-doped fiber amplifier (OptoCom Mod. OI LNPA; EDFA2
in Fig.3) with a low saturation power ($\simeq$ 30 $\mu$W at 1550 nm) that
maintains the average power level of the output optical signal at a constant
level ($\simeq$ 18 mW). In this way, the power levels transmitted through
the DB FBG, for the laser emission tuned either at Fabry-Perot resonances or
anti-resonances of the structure or outside the stopband, were comparable.
The transmitted pulse train was detected in the time domain by a fast
sampling oscilloscope  (Agilent Mod. 86100A), with a low jitter noise and an
impulsive response of $\simeq$ 15 ps; a portion of the sinusoidal RF signal
that drives the Mach-Zehnder modulator was used as an external trigger for
the oscilloscope, thus providing precise synchronism among successive
pulses.  Off-resonance tunneling was achieved by a careful tuning control of
the laser spectrum which was detected  by monitoring the reflected signal,
available at port three of the optical circulator, using both an optical
spectrum analyzer (Anritsu Mod. MS9710B) with a resolution of 0.07 nm, and a
plane-plane scanning Fabry-Perot interferometer (Burleigh Model RC1101R)
with a free-spectral range of $\simeq$50 GHz and a measured finesse of
$\sim$ 180, which permits to resolve the Fabry-Perot resonances of the DB FBG
structures. The reflectivity spectrum of the DB FBG was first measured by
the Fabry-Perot interferometer and recorded on a digital oscilloscope by
disconnecting the laser diode from the input port of EDFA1 and sending to
the DB FBG the broadband amplified-spontaneous emission
signal of the optical amplifier. This trace is then used as a reference to
tune the pulse spectrum at the center of the off-resonance plateau between
the two central Fabry-Perot resonances of the recorded DB FBG structure.\\ 
Figure 4 shows a typical trace [curve
(1)], averaged over 64 acquisitions, of the tunneled optical pulses under
off-resonance tuning condition, as measured on the sampling oscilloscope,
for the 42-mm separation DB FBG structure, and compared to the corresponding
trace [curve (2)] recorded when the laser was detuned apart by $\sim 200$
GHz, i.e. far away from the stopband of the DB FBG structure. A comparison
of the two traces clearly shows that tunneled pulses are almost undistorted with  a
peak pulse advancement of $\simeq$ 248 ps; repeated measurements showed that
the measured pulse peak advancement is accurate within $\simeq \pm$ 15 ps,
the main uncertainty in the measure being determined by the achievement of
the optimal tuning condition. We checked that propagation through EDFA2 does not 
introduce any appreciable pulse distortion nor any measurable time delay dependence on the 
amplification level. Time delay measurements were repeated for the
five DB FBG structures, and the experimental results are summarized in Fig.5
and compared with the theoretical predictions of tunneling time as given by 
Eqs.(3) and (4). The dashed line in the figure shows the theoretical  transit  time,
from input ($z=0$) to output ($z=2L_0+L$) planes, versus barrier separation
$L$ for pulses tuned far away from the bandgap of FBG; in this case, the
transit time is given merely by 
the time spent by a pulse traveling along the fiber for a distance $L+2L_0$
with a velocity $c_0 /n_0$. The solid line is, in turn, the expected transit
time for off-resonance tunneling of pulses, according to the phase time
analysis [see Eqs.(3) and (4)]], which shows that the transit time {\em does not
substantially increase} as the barrier separation is increased (generalized
Hartman effect). The points in the figure are obtained by subtracting to the
dashed curve the measured pulse peak advancements for the five DB FBGs,
thus providing 
an experimental estimate of the tunneling transit time. Notice that, within
the experimental errors, the agreement between measured and predicted
transit times is rather satisfactory. For each of the five DB FBG
structures, the measured transit times are Superluminal; it is remarkable
that, for the longest barrier separation used, the transit time leads to a
Superluminal velocity of about 5$c_0$, the largest one measured in
tunneling experiments at optical wavelengths \cite{note2}. These paradoxically small transit times do
not represent, however, a genuine violation of Einstein causality
\cite{Chiao97,Steinberg93,Spielmann94}, and may be qualitatively explained
as the result of two simultaneous effects that are a 
signature of the wave nature of the tunneling processes \cite{note3}. On the one hand,
following Refs.\cite{Olkh92,Steinberg93,Spielmann94}, peak pulse advancement occurs
at each of the two barriers as a result of a reshaping phenomenon in which
the trailing edge of the pulse is preferentially attenuated than the leading
one; on the other hand, the independence of tunneling time on the barrier
distance can be explained, following Ref.\cite{Olkhovsky01}, as an effective
``acceleration" of the forward traveling waves in the intermediate
classically-allowed region that arises in consequence of the destructive
interference between the two barriers.
\section{Conclusions}
In this paper  off-resonant tunneling of optical pulses has been
experimentally investigated in fiber Bragg photonic barriers. Tunneling time
measurements have been shown to be in good agreement with theoretical
predictions based on phase time analysis and have unambiguously confirmed
that, for opaque barriers, the tunneling time is
independent not only of the barrier width, as previously shown in
Ref.\cite{Spielmann94},  but even of the barrier separation. Our results extend to the 
optical region previous experimental achievements performed at microwaves
\cite{Enders93} and
may be of interest in the field of optical tunneling and related issue of Superluminal 
propagation.\\  
\acknowledgments
The authors acknowledge M. Marano for his help in the experimental
measurements, and F. Fontana and G. Salesi for many useful discussions.

\newpage
\begin{table}
\caption{ Analogies between tunneling of optical waves and electrons in a
symmetric rectangular DB potential.}

\

\begin{tabular}{lll}
Photons &  Electrons \\
\tableline
 Equations & \\
$ \begin{array}{ccc}
du/dz & = &   i \delta u + i k_B V(z) v \\
dv/dz & = & -i \delta v -i k_B V(z) u \\
\end{array}
$ 
&  $ \frac{d^2 \psi}{dz^2}+ \frac{2 m}{\hbar^2} \left[ E-V(z) \right] \psi =
0 $ \\
\tableline
DB Transmission\tablenote{ \scriptsize 
For electrons, calculations are made assuming a mean-energy of incident
wavepacket equal to half of the barrier height, i.e. $E=V_0/2$, and assuming
off-resonance tunneling, i.e.  $\chi L$ is an integer multiple of $\pi$,
where  $\chi \equiv \sqrt{m V_0} / \hbar$ is the wavenumber of oscillatory
wavefunction  between the two barriers. $v_g \equiv \hbar \chi /m$ is the
group-velocity of free wavepacket.} (off-resonance) & \\
 $ T=|t|^2= 1/ {\rm cosh}^2 (2 k_B V_0 L_0) $ & $T=|t|^2= 1/ {\rm cosh}^2 (2
\chi L_0) $ 
&
 \\
\tableline
Phase time$^{\rm a}$ (off-resonance) \\
$\tau={\rm Im} \left\{  \frac{\partial {\rm ln}(t)}{\partial \omega}\right\}
=\tau_1+\tau_2$ & $\tau= \hbar \; {\rm Im} \left\{  \frac{\partial {\rm
ln}(t)}{\partial E}\right\} =\tau_1+\tau_2$ &\\
$\tau_1=[n_0/(c_0 k_B V_0)] {\rm tanh}(2k_BV_0L_0)$ & $\tau_1=[2/(\chi v_g)]
{\rm tanh}(2 \chi L_0)$ \\
$\tau_2=(n_0 L/c_0) / {\rm cosh}(2k_BV_0L_0)$ & $\tau_2=(L/v_g) / {\rm
cosh}(2 \chi L_0)$ 
\end{tabular}
\end{table}

\newpage
\begin{center}
{\bf Figure Captions.}
\end{center}
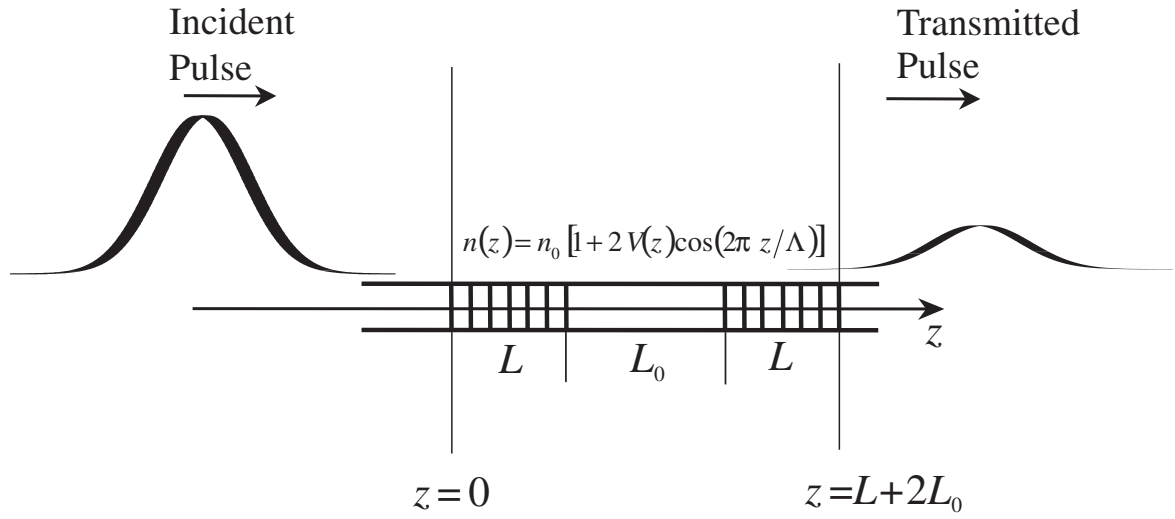
{\bf Fig.1.}
Schematic of tunneling through a rectangular DB photonic structure.\\
\\
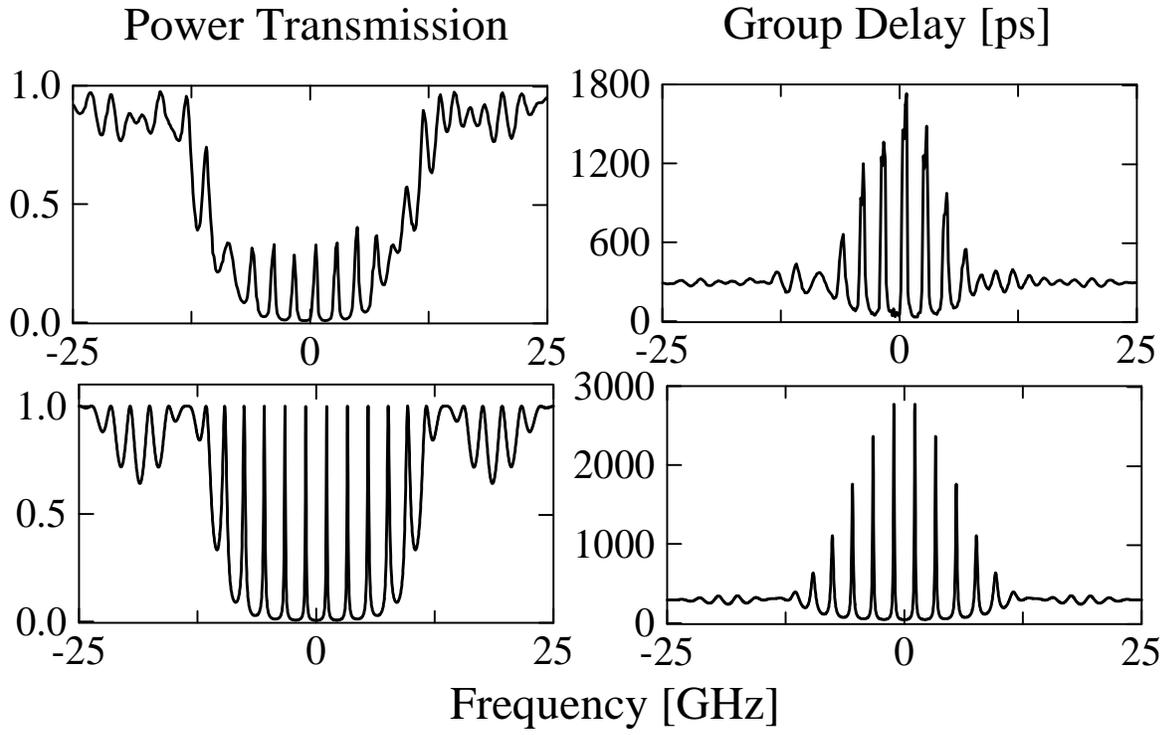
{\bf Fig.2.} 
Spectral power transmission (left) and group delay (right) for a DB FBG
structure for $L_0=8.5$ mm, $L=42$ mm, $V_0=0.9 \times 10^{-4}$,
$n_0=1.452$, and $\omega_B=1.261 \times 10^{15}$ rad/s.  Upper and lower
figures refer to measured and predicted spectral curves, respectively.\\
\\
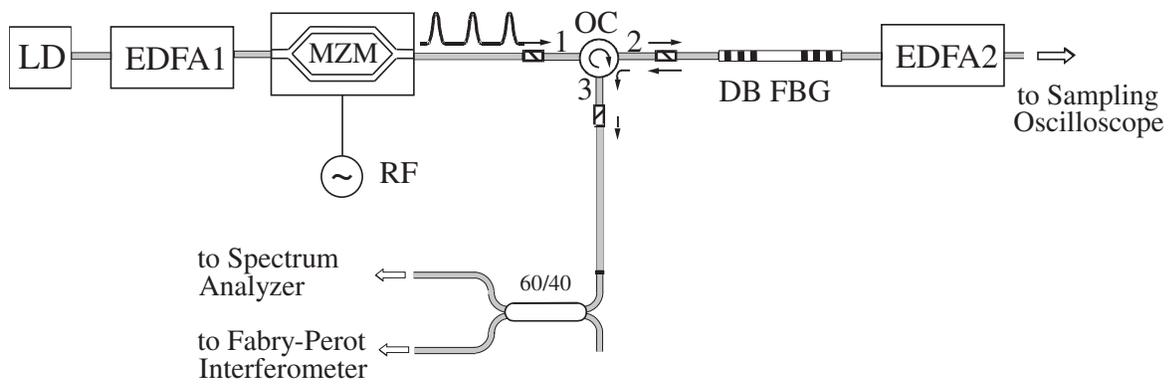
{\bf Fig.3.} 
Schematic of the experimental setup. LD: tunable laser diode; MZM:
Mach-Zehnder waveguide modulator; OC: optical circulator;
 EDFA1 and EDFA2: erbium-doped fiber amplifiers; RF: radio-frequency
synthetizer.\\
\\
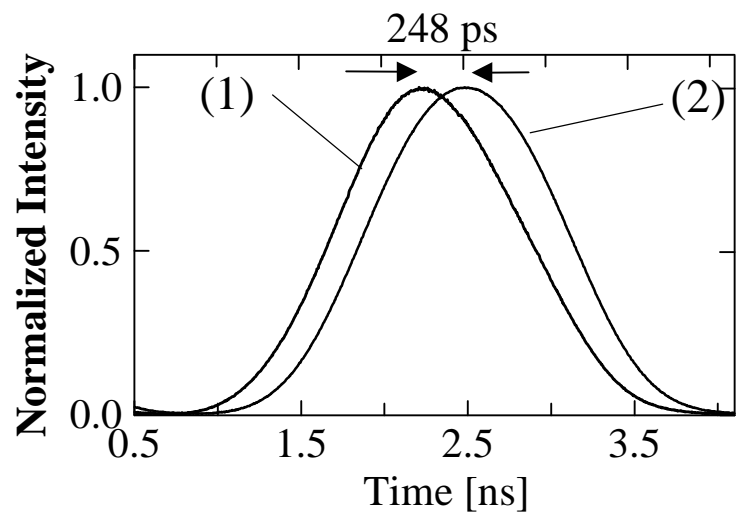
{\bf Fig.4.} Pulse traces recorded on the sampling oscilloscope
corresponding to the transmitted pulse
for off-resonance tunneling (curve 1) and reference pulse propagating
outside the stopband of the structure (curve 2) for the 42-mm separation DB
FBG.\\
\\
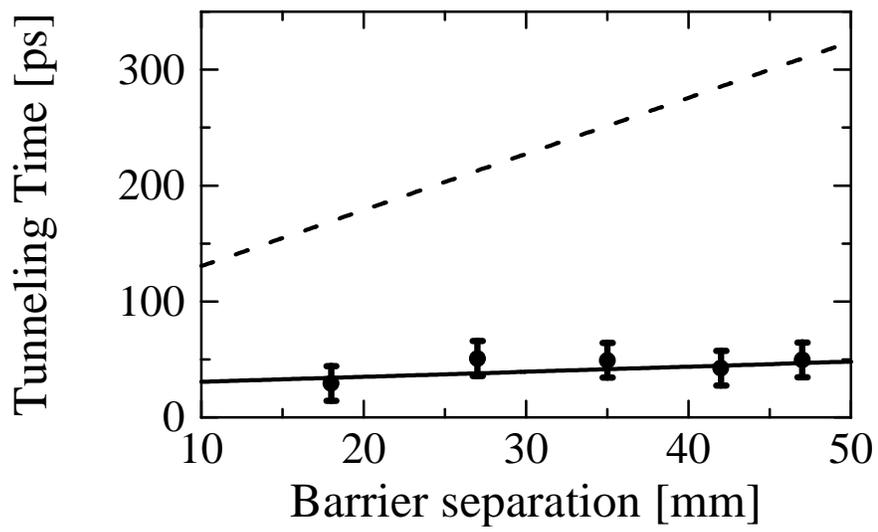
{\bf Fig.5.} Off-resonance tunneling time versus barrier separation $L$ for
a rectangular symmetric DB FBG structure. The solid line is the theoretical
prediction based on group delay calculations (Table I); dots are the
experimental points as obtained by time delay measurements; the dashed curve
is the transit time from input ($z=0$) to output ($z=L+2L_0$) planes for a
pulse tuned far away from the stopband of the FBGs.\\ 
%
%
\newpage
%
%
\newpage
%
%
\newpage
%
%
\newpage 
%
%

\begin{thebibliography}{99}

\bibitem{Hauge89}
E.H. Hauge and J.A. St\o vneng, Rev. Mod. Phys. {\bf 61}, 917 (1989).
\bibitem{Olkh92} V.S. Olkhovsky and E. Recami, Phys. Rep. {\bf 214}, 339 (1992);
V.S.Olkhovsky, E.Recami, F.Raciti and A.K.Zaichenko, {\em J. de Physique--I
(France)} {\bf 5}, 1351 (1995); V.S.Olkhovsky, E.Recami and J.Jakiel, ``Unified
time analysis of photon and nonrelativistic particle tunnelling", Lanl Archives
\# quant-ph/0102007.
\bibitem{Landauer94}
 R. Landauer and Th. Martin, Rev. Mod. Phys. {\bf 66}, 217 (1994).
\bibitem{Chiao97}
R.Y. Chiao and A.M. Steinberg, Prog. Opt. {\bf 37}, 345 (1997).
\bibitem{Nimtz97}
G. Nimtz and G.W. Heitmann, Progress in Quant. Electron. {\bf 21}, 81 (1997).
\bibitem{Hartman62}
T.E. Hartman, J. Appl. Phys. {\bf 33}, 3427 (1962).
\bibitem{Recami00}
E. Recami, F. Fontana and R. Garavaglia, Int. J. Mod. Phys. A {\bf 15},
2793 (2000), and refs. therein. For an extended review about Superluminal motions within special
relativity, see: E.Recami, Rivista N. Cim.  {\bf 9} (6), pp.1--178 (1986); Found. Phys. {\bf 31}, 1119 (2001).
\bibitem{Pablo00}
A.P. Barbero, H.E. Hernandez-Figueroa and E. Recami, Phys. Rev. E {\bf 62},
8628 (2000).
\bibitem{Enders93}
A. Enders and G. Nimtz, Phys. Rev. B {\bf 47}, 9605 (1993), and refs. therein;
G. Nimtz, A. Enders and H. Spieker, J. Phys.--I (France) {\bf 4}, 565 (1994);
H.M.Brodowsky, W.Heitmann and G.Nimtz, Phys. Lett. A {\bf 222}, 125 (1996).
For an earlier experiment on microwave tunneling see also: G. Nimtz and A.
Enders, J. Phys. I (France) {\bf 2}, 1693 (1992).
\bibitem{Ranfagni93} A.Ranfagni, P.Fabeni, G.P.Pazzi and D.Mugnai, Phys. Rev.
E {\bf 48}, 1453 (1993); D. Mugnai, A. Ranfagni and L. Ronchi, Phys. Lett.
A {\bf 247}, 281 (1998).
\bibitem{Steinberg93}
A.M. Steinberg, P.G. Kwiat and R.Y. Chiao, Phys. Rev. Lett. {\bf 71}, 708
(1993).
\bibitem{Spielmann94}
Ch. Spielmann, R. Szip\"{o}cs, A. Stingl and F. Krausz, Phys. Rev. Lett.
{\bf 73}, 2308 (1994).
\bibitem{Balcou97}
P. Balcou and L. Dutriaux, Phys. Rev. Lett. {\bf 78}, 851 (1997).
\bibitem{Merzbacher70}
See, for example, E. Merzbacher, {\em Quantum Mechanics} (Wiley, New York,
1970), Chaps. 6 and~7.
\bibitem{Capasso86}
F. Capasso, K. Mohammed and A.Y. Cho, IEEE J. Quant. Electron. {\bf QE-22},
1853 (1986).
\bibitem{Chiao91}
R.Y. Chiao, P.G. Kwiat and A.M. Steinberg, Physica B {\bf 175}, 257 (1991);
Th. Martin and R. Landauer, Phys. Rev. A {\bf 45}, 2611 (1992); J. Jakiel,
V.S. Olkhovsky and E. Recami, Phys. Lett. A {\bf 248}, 156 (1998).
\bibitem{Hupert77}
For early works on the analogy between electron and photon tunneling, see: 
T. Tsai and G. Thomas, Am. J. Phys. {\bf 44}, 636 (1976); J.J. Hupert, Am. J. Phys.
{\bf 44}, 636 (1977) and references therein.
\bibitem{Cuevas96}
E. Cuevas, V. Gasparian, M. Ortu\~{n}o and J. Ruiz, Zeit. Phys. B {\bf
100}, 595 (1996).
\bibitem{Olkhovsky01}
V.S. Olkhovsky, E. Recami and G. Salesi, ``Tunneling through two successive
barriers and the Hartman (Superluminal) effect", Lanl Archives \#
quant-ph/0002022.
\bibitem{Longhi01}
S. Longhi, Phys. Rev. E {\bf 64}, 037601 (2001).
\bibitem{Longhi01b}
S. Longhi, M. Marano, P. Laporta, and M. Belmonte, Phys. Rev. E {\bf 64}, 055602(R) (2001).
\bibitem{Headings63}
J. Headings, J. Atm. Terr. Phys. {\bf 25}, 519 (1963); P. Thanikasalam, R.
Venkatasubramanian, and M. Cahay, IEEE J. Quant. Electron. {\bf 29}, 2451 (1993); H. Yammamoto,
K. Miyamoto, and T. Hayashi, Phys. Stat. Sol. B {\bf 209}, 305 (1998).
\bibitem{note0}
The analogy between tunneling of electrons and photons in superlattice structures and
closed-form solutions for the tunneling times have been recently reported, in the
general case, by  P. Pereyra [P. Pereyra, Phys. Rev. Lett. {\bf 84}, 1772
(2000)]. The derivation of photon tunneling times presented here follows a
different and simpler approach, based on coupled-mode equations for Bragg
scattering, which is suited for FBG structures.
\bibitem{Ergodan97}
See, for instance, T. Ergodan, J. Lightwave Technol. {\bf 15}, 1277 (1997).
\bibitem{note1}
Among others, the phase time has been shown to be
consistent with tunneling time measurements in 
previous experiments on optical tunneling through photonic bandgaps
\cite{Steinberg93,Spielmann94}.
\bibitem{Ryu89}
S. Ryu, Y. Horiuchi and K. Mochizuky, J. Lightwave Technol. {\bf 7}, 1177
(1989).
\bibitem{note2}
We mention that Superluminal group velocities as large as $\sim 10 c_0$
were observed by G. Nimtz and coworkers, but in microwave transmission experiments.
\bibitem{note3}
We remark that the front velocity of a discontinuous step-wise signal that
propagates through the DB FBG structure is always equal to $c_0$, so there
is no violation of Einstein's causality in our experimental findings; this
circumstance follows technically from the analytic properties of spectral
transmission $t(\omega)$ and has been discussed in Ref.\cite{Longhi01b}; for
general discussions on this point see also: G. Nimtz, Eur. Phys. J. B {\bf 7}, 523 (1999);
M. Mojahedi, E. Schamiloglu, F. Hegeler, and K.J. Malloy, Phys. Rev. E {\bf 62},
5758 (2000) and references therein.

\end{thebibliography}
\end{document}